\documentstyle[12pt,epsfig]{article}

\begin {document}

\title{\bf CHARM HADROPRODUCTION IN $k_T$-FACTORIZATION
APPROACH}

\author{M.G.Ryskin and A.G.Shuvaev \\
Petersburg Nuclear Physics Institute, \\
Gatchina, St.Petersburg 188350 Russia \\
Yu.M.Shabelski \\
The Abdus Salam International Centre \\
for Theoretical Physics, Trieste, Italy \\
and \\
Petersburg Nuclear Physics Institute, \\
Gatchina, St.Petersburg 188350 Russia\footnote{Permanent
address}}
\date{} \maketitle

\vspace{0.5cm}

\begin{abstract}
We compare the theoretical status and the numerical predictions of two
approaches for heavy quark production in the high energy hadron collisions,
namely the conventional LO parton model with collinear approximation and
$k_T$-factorization approach. The main assumptions used in the calculations
are discussed. To extract the differences coming from the matrix elements
we use very simple gluon structure function and fixed coupling. It is shown
that the $k_T$-factorization approach calculated formally in LO and with
Sudakov form factor accounts for many contributions related usually to
NLO (and even NNLO) processes of the conventional parton model

\end{abstract}

\vspace{2cm}

E-mail RYSKIN@THD.PNPI.SPB.RU

E-mail SHABELSK@THD.PNPI.SPB.RU

E-mail SHUVAEV@THD.PNPI.SPB.RU

\newpage

\section{Introduction}

The investigation of the heavy quark production in high energy hadron
collisions provides a method for studying the internal structure
of hadrons. Realistic estimates of the cross section of the heavy quark
production, as well as their correlations, are necessary in order to plan
the experiments on existing and future accelerators.

QCD is the theory of quark and gluon interactions. The description of
hard hadron collisions is possible only at the phenomenological level,
reducing the hadron interactions to the parton-parton ones, treated
in terms of the hadron structure functions. The hard process cross
section is expressed as the convolution of the parton
distributions in the colliding hadrons with the cross section of the
elementary sub-process given by the square of the matrix element,
calculated in the perturbative QCD. Such an approach can be justified
only in the Leading Logarithmic Approximation (LLA) in the leading order
(LO), or in the next to leading order (NLO) and so on. Strictly
speeking, it is in disagreement with the quantum field
theory, where the amplitudes, rather than the probabilities, should be
added and/or multiplyed. Thus all theoretical approaches to the high
energy hadron interactions are in some sense phenomenological ones.

There are several phenomenological approaches to the calculations
of hard processes cross sections in hadronic reactions.

The most popular and technically simplest approach is the so-called
QCD collinear approximation, or parton model (PM). In this model all
particles involved are assumed to be on mass shell, having only
longitudinal components of the momenta, and the cross section is
averaged over two transverse polarizations of the incident gluons.
The virtualities $q^2$ of the initial partons are taken into account
only through their densities (structure functions). The latter are
calculated in LLA using DGLAP evolution equation and available
experimental data. The probabilistic picture of "frozen" partons
underlies this way of proceeding. The cross sections of QCD subprocess
are calculated usually in the LO, as well as in the NLO
\cite{1,2,NDE,Beer,Beer1}. However the transverse momenta of the incident
partons are neglected in the QCD matrix elements in the direct analogy
with the Weizsaecker-Williams approximation in QED. It allows to describe
quite reasonably the experimental data on the total cross sections and
one-particle distributions of produced heavy flavours, however it can not
\cite{BEAT,BEAT1} reproduce,
say, the azimuthal correlations \cite{MNR} of two heavy quarks, as well
as the distributions over the total transverse momentum of heavy quarks
pairs \cite{FMNR}, which are just determined by the transverse momenta
of incident partons.

The simplest way to incorporate the transverse momenta of the incident
partons for the proper description of heavy quark
correlations was suggested in \cite{FMNR}, as the random shift of these
momenta according to certain exponential distribution. However, this
procedure has no serious theoretical background. While the shift of the
order of $\Lambda_{QCD}$ ($\langle k_T \rangle \sim$ 300~MeV) looks to
be reasonable as having an origin in confinement forces at large distances,
the value $\langle k_T \rangle \sim$ 1~GeV or more should be explained
in terms of the perturbative QCD. Moreover, the averaged value of
the shift seems to be dependent on the initial energy, kinematical region,
etc. The simultaneous description of one-particle distributions over
the transverse and longitudinal momenta seems to be impossible \cite{Shab}.

Another natural method to account for the incident parton transverse
momenta is referred to as $k_T$-factorization \cite{CCH,CE,MW,CH,CC}, or
the theory of semihard interactions \cite{GLR,LR,8,lrs}. Here the Feynman
diagrams are evaluated for the virtual incoming partons with all possible
polarizations. Since there are no reason to neglect the gluon transverse
momenta $q_{1T}$ and $q_{2T}$ in comparison with the quark masses and
transverse momenta in the small $x$ domain, the QCD matrix elements of the
sub-processes become more complicated. We obtain them in the LO, the NLO
calculations being much more technically hard in this approach than in the
conventional PM. On the other hand, the essential and, most probably, the
major part of the NLO (and part of NNLO) corrections to the LO PM related
to the finite values of parton transverse momenta are already included
into LO contribution in case of $k_T$-factorization. Thus one can hope
that NLO, NNLO... corrections will not be too large in $k_T$-factorization,
making it to be more constructive.

The matrix element describing collision of the off-shell partons is not,
of course, gauge invariant. This is the shortness of $k_T$-factorization.
One can use it only in a "physical" (axial or planar) gauge, where the LO
Feynman graphs responsible for the parton (DGLAP or BFKL) evolution take
a ladder-like form, and the hard cross section can be factorized, that is
written as a convolution of the initial parton wave functions with
the hard sub-process matrix element squared. Recall, that an
unintegrated parton distribution in this case is given not just
by the derivative of the integrated parton density with respect to the
scale, but includes the double logarithmic (Sudakov) form factor
as well. This point has been discussed in more detail in
\cite{DDT} (see also Sect.~3 of the present paper). In Feynman
gauge this double logarithmic contribution comes from the graphs
where the soft gluons embrace the "hard" blob, as it is shown,
for example, in Fig.~1b (contrary to the PM case shown in Fig.~1a).

Note, that when the initial energy grows, the momentum fraction $x$ carrying
by the "active" (participating in hard interaction) parton decreases.
As a rule the smaller $x$ the larger number of steps the parton passes
in the evolution. Having emitted a lot of evolution gluons the parton gains a
large transverse momentum $k_T$. It explains why a large phenomenological
$k_T$ shift is needed to describe the high energy data in terms of the PM.

Today it is clear that the $k_T$-factorization approach is self-consistent
and predicts the results in reasonable agreement both with the experimental
data and with the collinear approximation when the last one can be applied.
However, the detailed numerical analysis of hard processes in the framework
of $k_T$-factorization is not provided yet.

The predictions of all phenomenological approaches depend significantly
on the quark and gluon structure functions. The last ones are more or
less known experimentally from the data of HERA, but unknown at very
small values of Bjorken variable $x < 10^{-4}$. However it is just the
region that dominates in the heavy quark production at high
energies\footnote{For example, in the case of charm production, $m_c$ =
1.4GeV, at LHC, $\sqrt{s}$ = 14 TeV, the product $x_1x_2$ of two gluons
(both $x_1$ and $x_2$ are the integral variables) is equal to
$4\cdot 10^{-8}$ and the applicapability of the existing structure functions
at so small $x$ is not clear, see discussion in \cite{ASS}.}.

The main goal of this paper is to compare the
results of the conventional parton model and the $k_T$-factorization
approach and to explain the source for the differences in their
predictions. We present the main formalism of these approaches as well as
the numerical results. To clarify the results we consider very high energies,
including the non-realistical one $\sqrt{s}$ = 1000 TeV, and use the
simple "toy" gluon structure function of the nucleon.

\section{Conventional parton model approach}

The conventional parton model expression for the heavy
quark hadroproduction cross sections has the factorized form
\cite{CSS}:
\begin{equation}
d\,\sigma (a b \rightarrow Q\overline{Q})\ =\
\sum_{ij} \int dx_i dx_j G_{a/i}(x_i,\mu_F) G_{b/j}(x_j,\mu_F)
d\,\hat{\sigma} (i j \rightarrow Q \overline{Q})\ ,
\label{pm}
\end{equation}
where $G_{a/i}(x_i,\mu_F)$ and $G_{b/j}(x_j,\mu_F)$ are the structure
functions of partons $i$ and $j$ in the colliding hadrons $a$ and $b$,
$\mu_F$ is the factorization scale (i.e. virtualities of incident
partons) and $d\,\hat{\sigma} (i j \rightarrow Q \overline{Q})$ is
the cross section of the subprocess calculated in the
perturbative QCD. It can be written as a sum of
LO and NLO contributions,
\begin{equation}
d\,\hat{\sigma} (i j \rightarrow Q\overline{Q})\ =\
\frac{\alpha_s^2(\mu_R)}{m_Q^2} \biggl( f^{(0)}_{ij} +
4 \pi \alpha_s(\mu_R) \Bigl[f^{(1)}_{ij} +
\bar{f}^{(1)}_{ij} \ln(\mu^2_R/M_Q^2) \Bigr] \biggr) \;,
\label{shat}
\end{equation}
where $\mu_R$ is the renormalization scale, and the functions 
$f^{(o)}_{ij}$, $f^{(1)}_{ij}$ as well as $\bar{f}^{(1)}_{ij}$ actually 
depend only on a single variable
\begin{equation}
\rho\ =\ \frac{4m_Q^2}{\hat{s}}\ , \qquad \hat s\ =\ x_i x_j
s_{ab}\ .
\end{equation}
The expression (\ref{pm}) corresponds
to the process shown schematically in Fig.~1a. The main
contribution to the heavy flavour production cross section at
small $x$ is known to come from gluons, $i = j = g$.

The principal uncertainties of any numerical QCD calculation are
the consequences of unknown scales \footnote{These uncertainties have
to disappear after all high order contributions are summed up. There
is the opinion that the strong (weak) scale dependence in LO or NLO
means the large (small) contribution of high order diagrams, but it
is not the case. The strong or weak scale dependence in LO or NLO
indicates only the same for the higher orders, which nevertheless can
be numerically small or large at some particular fixed scale.}, $\mu_F$
and $\mu_R$ and the heavy quark mass, $m_Q$. Both scales (sometimes they
are assumed to be equal) are to be of the order of hardness of the
treated process, however which value is better to take,
$m_Q$, $m_T = \sqrt{m_Q^2 + p_T^2}$ or $\hat{s}$, remains to be
unknown. Principally, the uncertainties should not be essential because
of the logarithmical dependence on these parameters.
Unfortunately the masses of $c$ and $b$ quarks are not large enough
and it leads to numerically large uncertainties (see e.g. \cite{FMNR})
at the existing energies for fixed target experiments.

Another principal problem of parton model is the collinear
approximation. The transverse momenta of the incident partons,
$q_{iT}$ and $q_{jT}$ are assumed to be zero, as is shown in Fig.~1a,
and their virtualities enter only via structure
functions; the cross section 
$d\,\hat{\sigma} (i j \rightarrow Q\overline{Q})$ is assumed to be 
independent on the virtualities. This approximation considerably simplifies 
the calculations.

The conventional NLO parton model approach with collinear approximati\-on
works quite reasonably for one-particle distributions and the total
cross sections; at the same time there is the serious disagreement with
the data on azimuthal correlations and on the transverse momentum
distributions of the heavy-quark pair. The reason is quite clear.
Consider the processes of the fusion of two partons, having transverse
momenta $q_{1T}$ and $q_{2T}$, into $Q\overline{Q}$ pair with transverse
momenta $p_{1T}$ and $p_{2T}$. In LO parton model
$q_{1T} + q_{2T} = p_{1T} + p_{2T} =0$, and the distribution on the
transverse momenta of the quark pair coinsides with the distribution on the
total transverse momentum of the initial partons. NLO gives a
correction, numerically not very large because the shapes of $p_T$
distribitions in LO and NLO are rather close to each other, see
\cite{NDE,MNR1}. In this case the qualitative features in NLO
should be the same as LO ones, so the final transverse
momentum distributions of the heavy-quark pair only slightly
differ from $\delta$-functions, being in contradiction with the
existing data \cite{BEAT,FMNR}.

It was shown \cite{FMNR}, that the conventional NLO parton model
can describe the experimental data on azimuthal correlations
and the momentum distributions of the heavy-quark pair, assuming the
comparatively large intrinsic transverse momentum of incoming partons
($k_T$ kick) about 1~GeV/c. This procedure is, to a large extent,
arbitrary. Let $\vec{p}_T (Q\overline{Q})$ be the total transverse
momentum of the pair. For each event the heavy quark pair,
$Q\overline{Q}$, is boosted firstly to the rest in the longitudinal
direction. Then a second transverse boost is performed, giving the
pair a transverse momentum $p_T(Q\overline{Q}) = k_{1,T} + k_{2,T}$;
$k_{1,T}$ and $k_{2,T}$ are the transverse momenta of the
incoming partons, which are chosen randomly, with their moduli
distributed according to
\begin{equation}
\frac1N\ \frac{dN}{dk_T^2}\ =\ \frac{1}{\langle k_T^2 \rangle}
\exp(-k_T^2/\langle k_T^2 \rangle)\ .
\label{kd}
\end{equation}
Alternatively, one can proceed as in the previous case, but giving the
additional transverse momentum $k_{1,T}+k_{2,T}$ to the whole
final-state system (at the NLO in QCD, this means the $Q\overline{Q}$
pair plus a light parton), and not to the $Q\overline{Q}$ pair only.
These two methods lead to similar results \cite{FMNR}.

However the $k_T$ kick method has no good theoretical background.
The general phenomenological expression for the heavy quark
production can be written\footnote{We omit here for the simplicity all
not important factors.} as a convolution of the initial
transverse momenta distributions, $I(q_{1T})$ and $I(q_{2T})$,
with squared modulo of the matrix element,
\begin{equation} \sigma_{QCD}(Q\overline{Q})\ \propto\
\int d^2 q_{1T} d^2 q_{2T} I(q_{1T}) I(q_{2T}) \vert M(q_{1T},
q_{2T}, p_{1T}, p_{2T}) \vert ^2 \;.
\label{ge}
\end{equation}

Here there are two possibilities: \\
i) the typical gluon transverse momenta are much
smaller than the transverse momenta of produced heavy quarks,
$q_{iT} \ll p_{iT}$, or \\
ii) all transverse momenta are of the same order,
$q_{iT} \sim p_{iT}$.

In the first case one can replace both initial distributions
$I(q_{iT})$ by $\delta$-functions. It reduces the expression (\ref{ge})
to the simple form (collinear approximation):
\begin{equation}
\sigma_{coll.}(Q\overline{Q})\ \propto\ \vert
M(0, 0, p_{1T}, p_{2T}) \vert ^2
\end{equation}
in total agreement with Weizsaecker-Williams approximation.

In the second case we can not a priory expect good results from the
Weizsaecker-Williams approximation. However it gives the reasonable
numerical results in some cases.

The $k_T$ kick \cite{FMNR} discussed above effectively accounts for the
transverse momenta of incident partons. It is based on the expression which
symbollically reads
\begin{equation}
\sigma_{kick}(Q\overline{Q})\ \propto\ I(q_{1T}) I(q_{2T})
\otimes \vert M(0, 0, p_{1T}, p_{2T}) \vert ^2\ .
\end{equation}
The main difference from the general QCD expression, Eq.~(\ref{ge}),
is that due to the absence of $q_{iT}$ in the matrix element the value
$\langle k_T^2\rangle$ in Eq.~(\ref{kd}) has to be differently taken
for different processes (say, for heavy flavour production with comparatively
small or large $p_T$). The reason is that the functions $I(q_{iT})$
in general QCD expression decrease for large $q^2_{iT}$ as a weak power
(see next Sect.), i.e. comparatively slowly, and
the $q^2_{iT}$ dependence of the matrix element is more important.

%In the last one the corrections of the order of $q^2_{iT}/\mu^2$,
%where $\mu^2$ is the QCD scale, are small enough when
%$q^2_{iT}/\mu^2 << 1$ and they start to suppress a matrix element
%value when $q^2_{iT}/\mu^2 \sim 1$. So the essential values of the
%$q^2_{iT}$ depend on the hardness of the process.

\section{Heavy quark production in \newline $\bf
k_T$-factorization approach}

Consider another approach, when the transverse momenta of the
incident gluons in the small-$x$ region result from the diffusion
in the parton (gluon) evolution. The diffusion is described by the function
$\varphi(x,q^2)$ giving the gluon distribution at the fixed fraction
of the longitudinal momentum of the initial hadron, $x$, and the gluon
virtuality, $q^2$. It is approximately determined by the derivative of
the usual structure function. Although generally it is the function
of three variables, $x$, $q_T$ and $q^2$, the transverse momentum dependence
is comparatively weak since $q_T^2 \approx - q^2$ for small $x$ in LLA
in agreement with $q^2$-dependences of structure function.
Note that due to QCD
scaling violation the value $\varphi(x,q^2)$ for the realistic
structure functions increases with decreasing of $x$, and
$q_T$ becomes important in the numerical calculations.

The exact expression for $q_T$ gluon distribution can be obtained as a
solution of the evolution equation which, contrary to the parton model
case, is nonlinear due to interactions between the partons in small
$x$ region. The calculations \cite{Blu} of the $q_T$ gluon distribution
in leading order using the BFKL theory \cite{BFKL} result in difference
from our $\varphi(x,q^2)$ function only about 10-15\%.

Here we deal with Eq.~(\ref{ge}) with the matrix
element accounting for the gluon virtualities and polarizations. Since
it is much more complicated than in the parton model we
consider only LO of the subprocess $gg \to Q\bar{Q}$
which gives the main contribution to the heavy quark production cross
section at small $x$, see the diagrams in Fig. 2.
The lower and upper ladder blocks present the two-dimensional gluon
functions $\varphi(x_1,q_1^2)$ and $\varphi(x_2,q_2^2)$. In terms
of the conventional DGLAP parton distributions they can be determined
as \cite{GLR}
\begin{equation}
\label{xG}
\varphi (x,q^2)\ =\ 4\sqrt2\,\pi^3 \frac{\partial[xG(x,q^2)]}
{\partial q^2}\ .
\end{equation}
Such a definition of $\varphi(x,q^2)$ makes it possible to treat correctly
the effects arising from the gluon virtualities.

Strictly speaking we have to include double logarithmic Sudakov-like
form factor under the derivative in Eq.~(8) \cite{DDT}.
 It was shown in \cite{DDT} that the probability to find a parton
with a given longitudinal momentum fraction $x$ and transverse
momentum $k_t$ within the LLA may be written as
\begin{equation}
\label{b}
f_a(x,k^2_t)\ =\ \frac\partial{\partial k^2_t}\left[xa(x,k^2_t)
T_a(k^2_t,\mu^2)\right]\ ,
\end{equation}
where $a=g,q$; $T$ is the Double Logarithmic (DL) Sudakov form
factor and
\begin{equation}
\varphi_a(x,q^2)\ =\ 4\sqrt2\ \pi^3 f_a(x,q^2)\ .
\end{equation}

The first factor in (9) is evident. It corresponds to the real
contribution/emission in the DGLAP evolution. Changing the scale
from $\mu^2$ to $\mu^2+\delta\mu^2$ one produces a new parton
with $k^2_t$ between $\mu^2$ and $\mu^2+\delta\mu^2$. So
\begin{equation}
f_a(x,k^2_t)\ =\ \frac\partial{\partial k^2_t}[xa(x,k^2_t)]\ .
\end{equation}
Next factor
\begin{equation}
\label{Sud}
T\ =\ \exp\left(-C\int\limits^{\mu^2}_{k^2_t}
\frac{\alpha_s(q^2)}{2\pi}\ln\left(\frac{\mu^2}{q^2}\right)
\frac{dq^2}{q^2}\right)
\end{equation}
($C=C_F=(N^2_c-1)/2N_c$ for the quarks and $C=C_A=N_c$ for the
gluons) accounts for the virtual loop DGLAP contribution, which
is needed to provide an appropriate normalization of the parton
wave function and to satisfy the sum rules (i.e. the
conservation of momentum, flavour, etc.). $T$ plays the role of
survival probability, i.e. the probability not to emit an extra
partons (gluons) with the transverse momenta
$q'_t\subset[k_t,\mu]$.

Note that in DGLAP equation, written for the integrated
(including all $k_t\le\mu$) partons, there was a cancellation
between the real and virtual soft gluon DL contributions.
Emitting a soft gluon with the momentum fraction $(1-z)\to0$ one
does not change the $x$-distribution of parent partons. Thus the
virtual and real contributions originated by the $1/(1-z)$
singularity of a splitting function $P(z)$ cancels each other.

Contrary, in unintegrated case the emission of soft gluon (with
$q'_t>k_t)$ alters the transverse momentum of parent
($t$-channel) parton. Eq. (9) includes for this effect through
the derivative $\partial T(k^2_t,\mu^2)/\partial k^2_t$.

Thanks to this last derivative one obtains a positive
unintegrated distribution $f_a(x,k^2_t)$ even at rather large
$x$, where (due to the virtual term in DGLAP) the integrated
parton density $a(x,\mu^2)$ decreases with $\mu^2$.

Unfortunately, the factor $T$ is known with the DLog accuracy
only. When $k_t>\mu$ there are no double logarithms at all. So for 
$k_t>\mu$ we put $T=1$.

In terms of Feynman diagrams in a physical (axial) gauge the
factor $T$ comes from the self-\-energy graphs. In Feynman gauge
the self-\-energy contribution has no double logarithms. Such a
double logarithms come from the graphs with the gluons which
embrace the hard blob (an example is shown in Fig.1b). Any
diagram with a soft gluon emitted from one external line of hard
blob (say, gluon $q_2$) and absorbed on another external line
has the same double logarithm, except of the colour factor. The
sum of colour factors, corresponding to all three graphs (soft
gluon is absorbed by the heavy quark, antiquark or gluon $q_1$),
is equal to the colour factor for self-\-energy diagram. This
way one obtains in Feynman gauge the same expression (9).

Thus the true probability
to find a parton with the momentum fraction $x$ and transverse momentum
$k_T^2 = k^2$ is given by the Eq.~(9). (see \cite{DDT} for more
detail).  To simplify the discussion we postpone the more exact
Eq.~(9) to the end of Section 4 and start with the
Eq.~(8) omitting the double logarithmic factor $T$.

A different way to introduce the gluon $q_T$ distributions is suggested
in Ref.~\cite{KS} and based on the Fourier transform of the structure
functions.

In what follows we use Sudakov decomposition for quark momenta
$p_{1,2}$ through the momenta of colliding hadrons  $p_A$ and
$p_B\,\, (p^2_A = p^2_B \simeq 0)$  and transverse ones $p_{1,2T}$:
\begin{equation}
\label{1}
p_{1,2} = x_{1,2} p_B + y_{1,2} p_A + p_{1,2T}.
\end{equation}
The differential cross section of heavy quarks hadroproduction has the
form:\footnote{We put the argument of $\alpha_S$ to be equal to gluon
virtuality, which is very close to the BLM scheme\cite{blm}; (see also
\cite{lrs}).}
\begin{eqnarray}
\frac{d\sigma_{pp}}{dy^*_1 dy^*_2 d^2 p_{1T}d^2
p_{2T}}\,&=&\,\frac{1}{(2\pi)^8}
\frac{1}{(s)^2}\int\,d^2 q_{1T} d^2 q_{2T} \delta (q_{1T} +
q_{2T} - p_{1T} - p_{2T}) \nonumber \\
\label{spp}
&\times &\,\frac{\alpha_s(q^2_1)}{q_1^2} \frac{\alpha_s (q^2_2)}{q^2_2}
\varphi(q^2_1,y)\varphi (q^2_2, x)\vert M_{QQ}\vert^2.
\end{eqnarray}
Here $s = 2p_A p_B\,\,$, $q_{1,2T}$ are the gluon transverse momenta
and $y^*_{1,2}$  are the quark rapidities in the hadron-hadron c.m.s.
frame,
\begin{equation}
\label{xy}
\begin{array}{lcl}
x_1=\,\frac{m_{1T}}{\sqrt{s}}\, e^{-y^*_1}, &
x_2=\,\frac{m_{2T}}{\sqrt{s}}\, e^{-y^*_2},  &  x=x_1 + x_2\\
y_1=\, \frac{m_{1T}}{\sqrt{s}}\, e^{y^*_1}, &  y_2 =
\frac{m_{2T}}{\sqrt{s}}\, e^{y^*_2},  &  y=y_1 + y_2 \\
&m_{1,2T}^2 = m_Q^2 + p_{1,2T}^2. &
\end{array}
\end{equation}
$\vert M_{QQ}\vert^2$ is the square of the matrix element for the heavy
quark pair hadropro\-duction.

In LLA kinematic
\begin{equation}
\label{q1q2}
\begin{array}{crl}
q_1 \simeq \,yp_A + q_{1T}, & q_2 \simeq \,xp_B + q_{2T}.
\end{array}
\end{equation}
so
\begin{equation}
\label{qt}
\begin{array}{crl}
q_1^2 \simeq \,- q_{1T}^2, & q_2^2 \simeq \,- q_{2T}^2.
\end{array}
\end{equation}
(The more accurate relations are $q_1^2 =- \frac{q_{1T}^2}{1-y}$,
$q_2^2 =- \frac{q_{2T}^2}{1-x}$ but we are working in the kinematics
where $x,y \sim 0$).

The matrix element $M$ is calculated in the Born order of QCD without
standart simplifications of the parton model. In the axial gauge
$p^\mu_B A_\mu = 0$ the gluon propagator takes the form
$D_{\mu\nu} (q) = d_{\mu\nu} (q)/q^2,$
\begin{equation}
\label{prop}
d_{\mu\nu}(q)\, =\, \delta_{\mu\nu} -\, (q^\mu p^\nu_B \, + \, q^\nu
p^\mu_B )/(p_B q) \;.
\end{equation}

For the gluons in $t-$channel the
main contribution comes from the so called "nonsense" polarization
$g^n_{\mu\nu}$, which can be picked out by decomposing the numerator
into longitudinal and transverse parts:
\begin{equation}
\label{trans}
\delta_{\mu\nu} (q)\ =\ 2(p^\mu_B p^\nu_A +\, p^\mu_A p^\nu_B )/s\,
+ \, \delta^T_{\mu\nu} \approx\, 2p^\mu_B p^\nu_A /s\,\equiv\,
g^n_{\mu\nu}\ .
\end{equation}
The other contributions are suppressed by the powers of $s$. It is easy
to check that in axial gauge (\ref{prop}) $p_B^\mu d_{\mu \nu}=0$ and
$p_A^\mu d_{\mu \nu} \simeq -q_T^ \nu/y$. Thus we effectively get
\begin{equation}
\label{trans1}
d_{\mu\nu} (q)\ \approx\ -\,2\, \frac{p^\mu_B q^\nu_T}{y\,s}\ .
\end{equation}
Another way to obtain the same result is to use the transversality condition.
Since the sum of the diagram Fig.~2a-c is gauge invariant the product
$q_{1\mu} M_\mu = 0$, here $M_\mu$ denotes the amplitude of the gluon $q_1$
and hadron $p_B$ interaction described by the lower part of
Fig.~2a--c.  Using the expansion (\ref{q1q2}) for the $q_1$
momentum we obtain
$$ p_A^\mu M_\mu\ \simeq\ -\frac{q_{1T}^\mu}y
M_\mu\ , $$
which leads to Eq.~(\ref{trans1}) or
\begin{equation} \label{trans2}
d_{\mu\nu} (q)\ \approx\ \,2\, \frac{q^\mu_T q^\nu_T}{xys}\ ,
\end{equation}
if we do such
a trick for the vector $p_B$ too \footnote{Having in mind this
trick one can say that the matrix element is gauge-invariant in
$k_T$-factorization approach. However the polarization vectors of incoming
gluons $q_1$ and $q_2$ are not arbitrary taken but fixed as $-q_{1T}^\mu/y$
and $-q_{2T}^\nu/x$, respectively (see \cite{GLR} for more detail).}.
Both these equations for
$d_{\mu\nu}$  can be used but for the form (\ref{trans1}) one has to
modify the gluon vertex slightly (to account for several ways of gluon
emission -- see ref. \cite{3}):
\begin{equation}
\label{geff}
\Gamma_{eff}^{\nu}\ =\
\frac{2}{xys}\ \left[(xys - q_{1T}^2)\,q_{1T}^{\nu} - q_{1T}^2
q_{2T}^{\nu} + 2x\,(q_{1T}q_{2T})\,p_B^{\nu}\right]\ .
\end{equation}
As a result the colliding gluons can be treated as aligned ones and
their polarization vectors are directed along the transverse momenta.
Ultimately, the nontrivial azimuthal correlations must arise between
the transverse momenta $p_{1T}$ and $p_{2T}$ of the heavy quarks.

From the formal point of view there is a danger to loose the gauge
invariance in dealing with the off mass shell gluons. Say, in the
covariant Feynman gauge the new graphs (similar to the "bremsstruhlung"
from the initial or final quark line, as it is shown in Fig.~2d) may
contribute in the central plato rapidity region. However this is not the
fact. Within the "semihard" accuracy, when the function $\varphi(x,q^2)$
collects the terms of the form $\alpha_s^k(\ln q^2)^n(\ln (1/x))^m$ with
$n+m\ge k$, the triple gluon vertex (\ref{geff}) includes effectively all the
leading logarithmic contributions of the Fig.~2d type \cite{BFKL,8}.
For instance, the upper part of the graph shown in Fig.~2d corresponds
in terms of the BFKL equation to the $t$-channel gluon reggeization.
Thus the final expression is gauge invariant (except a small,
non-logarithmic, $O(\alpha_s)$ corrections) \footnote{Taking the gluon
polarization vector as $-q_{1T}^\mu/y$ we can completely eliminate the
leading logarithm terms arising from Fig.~2d.}.

Although the situation considered here seems to be quite opposite to the
parton model there is a certain limit \cite{3}, in which our formulae
can be transformed into parton model ones, namely when the quark 
transverse momenta, $p_{1,2T}$, are much larger than the gluon ones, 
$q_{1,2T}$.

\section{Results of numerical calculations}

Eq.~(\ref{spp}) enables to calculate straightforwardly all distributions
concerning one-particle or pair production. To illustrate the difference
between our approach and the conventional parton model we present first
of all the results of calculations of charm production ($m_c$ = 1.4 GeV
\cite{Nar,BBB}) with high $p_T$ at three energies, $\sqrt{s}$ = 1~TeV,
10~TeV and 10$^3$~TeV and with the same value of
\begin{equation}
x_T\ =\ \frac{2 p_T}{\sqrt{s}}\ =\ 0.02\ ,
\end{equation}
i.e. $p_T$ = 10~GeV, 100~GeV and 10$^4$~Gev for the above energies.

We will use QCD scales $\mu_F = m_T = \sqrt{m^2_c + p^2_T}$ and
$\mu_R = m_c$, i.e. fixed coupling with $N_f$ = 3 and $\Lambda$ = 248
MeV.

However there exists a problem coming from the infrared region, because
the functions $\varphi (x,q^2_2)$ and $\varphi (y,q^2_1)$ are unknown
at the small values of $q^2_{1,2}$. Moreover, for the realistic gluon
structure functions the value $\varphi(x,q^2)$ is positive in the
small-$x$ region and negative in the large-$x$ region. The boundary
between two regions, where $\varphi (x,q^2)$ =0, depends on the $q^2$,
therefore the relative contributions of these regions is determined by
the characteristic scale of the reaction, say, by the transverse
momentum $p_T$.

To avoid this additional problem, we present the numerical
calculations both in the $k_T$-factorization approach and in the
LO parton model, using the "toy" gluon distribution
\begin{equation}
\label{toy}
xG(x,q^2)\ =\ (1-x)^5 \ln(q^2/Q_0^2)\ ,
\end{equation}
for $q^2 > Q_0^2$, and $xG(x,q^2) = 0$ for $q^2 < Q_0^2$, with
$Q_0^2$ = 1~GeV$^2$.

After the calculations of charm production cross sections using
Eqs. (1) and (8), (14) one can see, that the values $d \sigma /d
x_T$ at $x_T$ = 0.02 are about 4--5 times larger in the
$k_T$-factorization approach than in the LO parton model. Really
this difference should not be considered as very large because,
as it was discussed, an essential part of NLO and NNLO
corrections is already included in the $k_T$-factorization, and
it is known that the sum of LO and NLO contributions in the
parton model is about 2-3 times larger than the LO contribution
\cite{Liu}.

To show the most important values of the variables $q_{1,2T}$
in Eq.~(\ref{spp}), as well as the kinematical region producing the main
difference with the conventional parton model, we plot by dots in Fig.~3
the results of the $k_T$-factorization approach with the restrictions
$\vert q_{1,2T} \vert \leq q_{max}$, as a function of $q_{max}$. The
running coupling is fixed as $\alpha_S(m^2_c)$ instead of
$\alpha_S(q^2_{1,2})$ in Eq.~(\ref{spp}). The dashed lines in Fig.~3
are the conventional parton model predictions with QCD scales
$\mu_F = \sqrt{m_c^2 + p_T^2}$, and $\mu_R = m_c$. One can see that
the $k_T$-factorization predictions exceed the parton model results
when $q_{max} \geq p_T$.

Let us check now that the $k_T$-factorization predictions really
coincide with the parton model ones, when the quark momenta,
$p_{1,2T}$, are much larger than the gluon ones, $q_{1,2T}$ \cite{3}.
However we have to compare them at the same values of structure functions.
When we submit Eq.~(\ref{spp}) to the conditions
$\vert q_{1,2T} \vert \leq q_{max}$ and neglect the $q_{1,2T}$ dependence
in the matrix element, we recover the parton model approximation, and get
the result proportional to $xG(x,q^2_{max}) \cdot yG(y,q^2_{max})$
due to the direct consequence of Eq.~(\ref{xG}) \cite{Kwi}
\begin{equation}
xG(x,q^2)\ =\ xG(x,Q_0^2) + \frac{1}{4\sqrt{2} \pi^3}
\int_{Q_0^2}^{q^2} \varphi(x,q_1^2) dq_1^2\ .
\end{equation}

At the same time the original parton model yields the result
proportional to $xG(x,\mu_F^2) \cdot yG(y,\mu_F^2)$ with
$\mu_F = \sqrt{m_c^2 + p_T^2}$. Hence we expect the parton model
to coincide with our calculations for the gluon distribution
Eq.~(\ref{toy}), $p_T \gg m_c$ and $\vert q_{1,2T} \vert \leq q_{max}$
after multiplying by an appropriate factor:
\begin{equation}
\label{f}
\frac{d \sigma}{d x_T} \left \vert _{q_{iT} \ll p_{iT}}\ =\
\frac{d \sigma}{d x_T} \right \vert _{PM}
\left (\frac {\ln(q^2_{max}/Q_0^2)}{\ln(p_T^2/Q_0^2)} \right )^2  \;.
\end{equation}

The right-hand side of Eq.~(\ref{f}) presented in Fig.~3 by solid curves
shows a good agreement with $k_T$-factorization approach (open dots) 
even when $q_{max}$ is only slightly smaller (at the highest energy) 
than $p_T$. The same values $d \sigma / d x_T$, as in Fig.~3, but 
differentiated with respect to $\ln{q_{max}}$ are presented in Fig.~4 
for $k_T$-factorization approach. It exhibits, especially at the high 
energies, the logarithmic
growth with $q_{max}$, until the value $q_{max} \sim p_T$. There is the
narrow peak in this region, which provides about 70-80 \% of the
$d \sigma / d x_T$ cross section integrated over the whole $q_{1,2T}$ phase
space. Its physical nature seems to be quite clear. There are two kinematical
regions for $t$-channel and $u$-channel diagrams, Fig.~2~a,b, giving
comparatively large contributions to $d \sigma / d x_T$ for the high
energies and high $p_T$ heavy quark production. One of them comes from the
conventional parton model kinematics when both initial gluon transverse
momenta, $q_{1,2T}$, are very small compared to $p_{1,2T}$, see Fig.~5a.
Another large contribution appears from the region where, say,
$q_{1T} \sim p_{1T}$, whereas $q_{2T}$ and $p_{2T}$ are
comparatively small, as it is shown in Fig.~5b. In this case the
quark propagator, 
$1/(\hat{p_1} - \hat{q_1}) - m_Q) = 1/(\hat{q_2} - \hat{p_2}) - m_Q)$, 
is close to the mass shell and gives rise to the narrow peak shown in
Fig.~4. The general smallness of such processes comes from
high-virtuality gluon propagator in Fig.~5b, and it is of the
same order as in the case of Fig.~5a, even with some suppression
of the diagram Fig.~5a, where the contribution from the region
of large rapidity difference of two produced heavy quark is
suppressed by the fermion propagator.

The diagram shown in Fig.~5b can be considered \cite{1} as one of the most
important contribution to the NLO parton model in the high energy limit,
because it have spin-one gluon in the $t$-channel. Due to all these
factors, combinatorics and the interference between diagrams, the summary
contribution of the peak to the total $d \sigma / d x_T$ value is several
times larger than the LO parton model contribution. In the Table 1 we
present the calculated ratios of the total heavy quark pair production
cross section, $R_{tot}$, and the ratio of $d \sigma /d x_T$, $R(x_T)$ at
$x_T$ = 0.02 (integrated over rapidity).

\begin{table}
\caption{The ratios of $c\bar c$ pair production in
$k_T$-factorization approach and in LO parton model}

\begin{center}
\vskip 20 pt
\begin{tabular}{||c|r|r|r|r|r||} \hline\hline

$\sqrt s$, TeV & 0.3 & 1   & 10  & 100 & 1000 \\

\hline

$R_{tot}$       & 4.0 & 4.0 & 4.0 & 3.9 & 3.9 \\

$R(x_T = 0.02)$  & 3.4 & 4.5 & 5.5 & 5.4 & 5.2 \\
\hline\hline
\end{tabular}
\end{center} \end{table}
\vskip 10 pt

One can see that the relative contribution of the discussed peaks
firstly increases due to increase the phase space. At the energy high
enough ($\sqrt{s} \sim 10$ TeV this contribution is saturated. After
that the relative contribution of the LO parton model increases
logarithmically with $p_T$, and have to dominate at the extremely high
energies and $p_T$, in academical asymptotic.

These results are confirmed by those presented in Fig.~6, where we
reproduce the data from Fig.~3 for LO parton model and
$k_T$-factorization approach with the condition
$\vert q_{1,2T} \vert \leq q_{max}$. For comparison we show by stars
the $k_T$-factorization predictions for all values $q_{1T}$ with the
only condition $\vert q_{2T} \vert \leq q_{max}$, versus $q_{max}$. The
last results are above the LO parton model even at not too large
$q_{2T}$ because the peaks, discussed above, are included into the integral
over $q_{1T}$.

Let us note that the calculation of $d \sigma /d x_T$ at $\sqrt{s}$ = 10 TeV,
$x_T$ = 0.02, and with restriction $\vert q_{1,2T} \vert \geq p_T/2$ 
(see Fig.~5c) gives only about 2\% of the total value of $d \sigma /d x_T$.

It is necessary to note that the essential values of $q_{1,2T}$ increase
in our calculations when the transverse momentum, $p_T$ of the
detected $c$-quark increases. At initial energy high enough, 
$q_{1,2T} \sim p_T$. In the $k_T$ kick language it means that the
$\langle k^2_T\rangle$ value also increases.

The values $d\sigma /dp_T$ are presented in Fig.~7. Here the lower dashed
curves correspond to the LO parton model, solid curves --- to
the $k_T$-factorization approach with unintegrated gluon function
$\varphi(x,q^2)$ given by the simplified expression (8) while the
dottted and dash-dotted curves embody the double-logarithmic $T$-form 
factor (12) and use Eqs.~(9) and (10) (see the discussion in section
3).  Recall that from the diagram viewpoint the $T$-factor sums
up the diagrams containing in the Feynman gauge the gluon lines
embracing the hard blob as it is shown in Fig.~1b.  In the axial
gauge the corresponding Double Log contribution comes from the
parton (gluon) self-energy insertion, i.e. from the term
proportional to $\delta(1-z)$ in the DGLAP evolution equation.
The $T$-factor reproduces a large part of the virtual NLO
corrections with respect to the conventional PM and diminishes
the cross section. It suppresses the contribution coming from
small values of parton virtualities, $q_{1,2}^2 \ll m_T^2 =
m_Q^2+p_T^2$ in Eq.~(\ref{spp}). Such a suppression becomes more
important at very large $p_T$ as in this case the PM
contribution, which is collected from the logarithmically large
interval of $q_{1,2}^2 \ll p_T^2$ $(\int_{Q_0^2}^{p_T^2}
dq_{1,2}^2/q_{1,2}^2)$, starts to dominate.

Recall that before we have no "scale" problem within the
$k_T$-factorization approach since we deal with the explicit
integral (\ref{spp}) over the parton virtualities $q_{1,2}^2$.
The dependence of QCD matrix element
$M(q_{1T},q_{2T},p_{1T},p_{2T})$ on the $q_{1,2T}$ values provides the
convergency to the integral "automatically" fixing the "scale" $\mu_F$.

After $T$-factor is "switched on" one gets the problem of scale back.
The scale $\mu^2$ is not fixed in (\ref{Sud}) at LO accuracy. The reasonable
$\mu^2$ values are expected to be somewhere between $m_T^2$ and
$\hat s =M_{Q \overline Q}^2 = xys$. The numerical difference between
these values is very small, as one can see from Fig.7.

The rapidity distributions of prodiced charm quark is presented in
Fig.~8. Here again the dashed curves are the results of LO parton model
calculations, the solid curves show the results of the
$k_T$-factorization approach with unintegrated gluon function
$\varphi(x,q^2)$ given by the simplified expression (8),
while the dotted curves embody the double-logarithmic $T$-form factor
(\ref{Sud}) with $T(m_T^2)$ and use Eqs.~(9) and (10). One can
see that the last form factor decrease the cross section
calculated in the $k_T$-factorization approach about 1.5 times.

\section{Conclusion}

We have compared the conventional LO Parton Model (PM) and the
$k_T$-factorization approach for heavy quark hadroproduction. In order to
concentrate on the specific features of the approach and not to deal with
the pecularities of the parton distributions a simple gluon density
(\ref{toy}) has been chosen.

It has been shown that the contribution from the domain with strong $q_T$
ordering ($q_{1,2T} \ll m_T=\sqrt{m_Q^2+p_T^2}$) coincides in
$k_T$-factorization approach with the LO PM prediction. However a very
numerically large contribution comes besides this in $k_T$-factorization
from the region $q_{1,2T}\ge m_T$.
It is kinematically related to the events
where the transverse momentum of heavy quark $Q$ is balanced not by the
momentum of antiquark $\overline Q$ but by the momentum of the nearest
gluon.

This configuration is associated with NLO (or even NNLO, if both
$q_{1,2T} \ge m_T$) corrections in terms of the PM
with fixed number of
flavours, i.e. without the heavy quarks in the evolution. Indeed, as it
was mentioned in \cite{1}, up to 80\% of the whole NLO cross section
originates from the events where the heavy quark transverse momentum is
balanced by the nearest gluon jet. Thus the large "NLO" contribution,
especially at large $p_T$, is explained by the fact that the virtuality
of the $t$-channel (or $u$-channel) quark becomes small in the region
$q_T \simeq p_T$ and the singularity of the quark propagator
$1/(\hat{p} - \hat{q}) - m_Q)$ in the "hard" QCD matrix element,
$M(q_{1T},q_{2T},p_{1T},p_{2T})$, reveals itself.

Including the double logarithmic Sudakov-type form factor $T$ into the
definition of unintegrated parton density (\ref{b}) one takes into account
an important part of the virtual loop NLO (with respect to PM) corrections.
Thus we demonstrate that $k_T$-factorization collects already at LO the major
part of the contributions which play the role of the NLO (and even NNLO)
corrections to the conventional PM. Therefore we hope that the higher order
(in $\alpha_S$) correction to the $k_T$-factorization would be rather small.

Another advantage of this approach is that a non-zero transverse momentum
of $Q \overline Q$-system ($p_{1T}+p_{2T}=q_{1T}+q_{2T}$) is naturally
achieved in $k_T$-factorization. The typical value of this momentum
($k_T$-kick) depends on the parton structure functions/densities. It
increases with the initial energy ($k_T$-kick increases with the decreasing
of the momentum fractions $x,y$ carried by the incoming partons) and with
the transverse momenta of heavy quarks, $p_T$. Thus one gets a possibility
to describe a non-trivial azimuthal correlation without introducing a large
"phenomenological" intrinsic transverse momentum of the partons.

A more detail study of the heavy quark production (including
the correlations) in the $k_T$-factorization approach based on the modern
realistic parton distributions will be published elsewhere.

\subsection*{Acknowledgements}
The main part of the presented calculations were carried out in ICTP.
One of us (Y.M.S) is grateful to Prof. S.Randjbar-Daemi for providing
this possibility and to the staff for creating good working conditions.
We are grateful to E.M.Levin who participated at the early stage of
this activity and to Yu.L.Dokshitzer, G.P.Korchemsky amd M.N.Mangano
for discussions.\\
This work is supported by grants NATO OUTR.LG 971390, RFBR 98-02-17629
and by Volkswagen Stiftung.

\newpage

{\bf Figure captions}

\vspace{.5cm}

Fig. 1. Heavy quark production in the LO parton model (a) and in
$k_T$-factorization approach with Sudakov T-factor (see discussions in
Sect. 3).

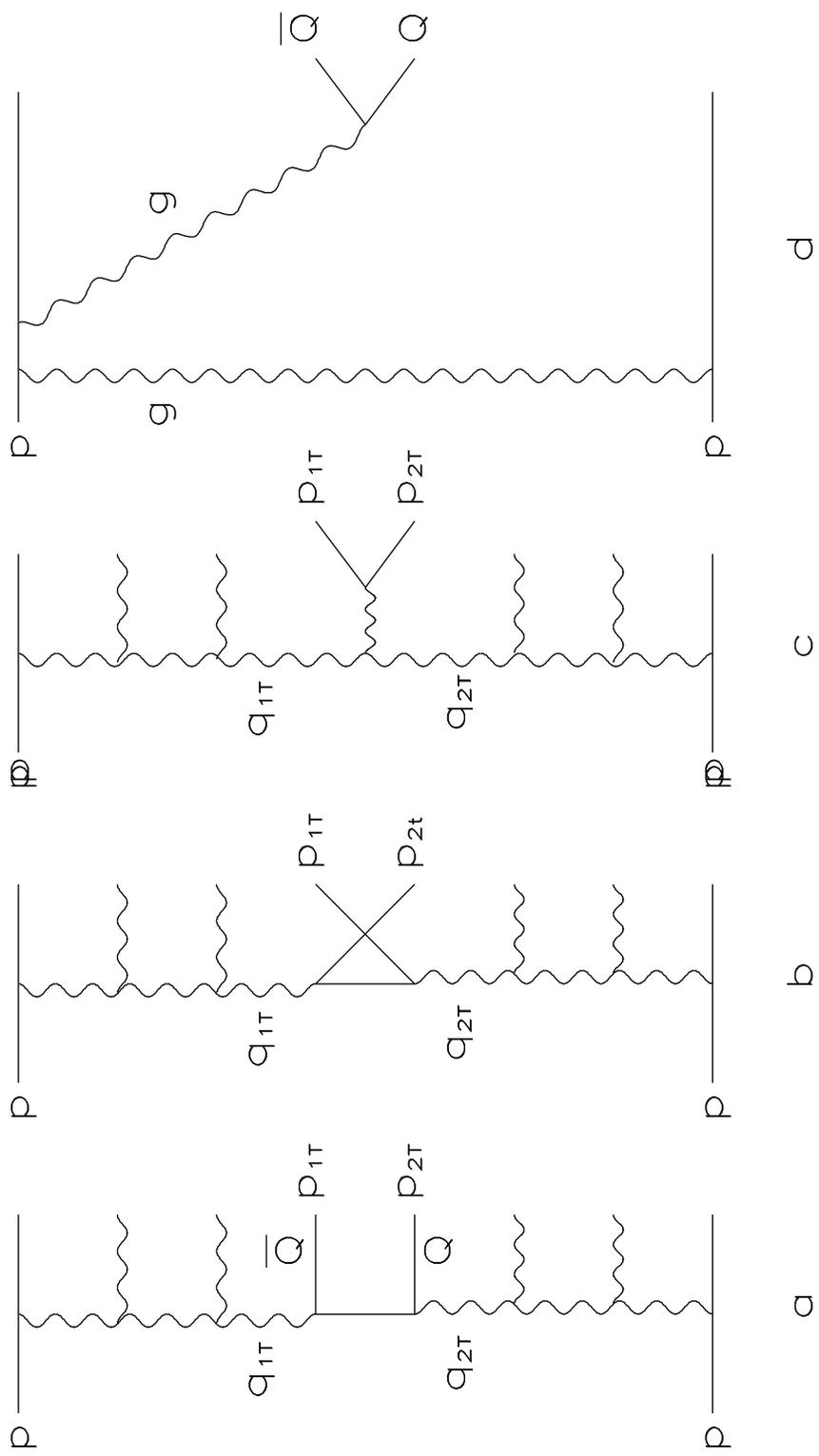
Fig. 2. Low order QCD diagrams for heavy quark production in $pp$
($p\overline{p}$) collisions via gluon-gluon fusion (a-c) and the
diagram (d) formally violating the factorization, that is restored
within logarithmic accuracy.

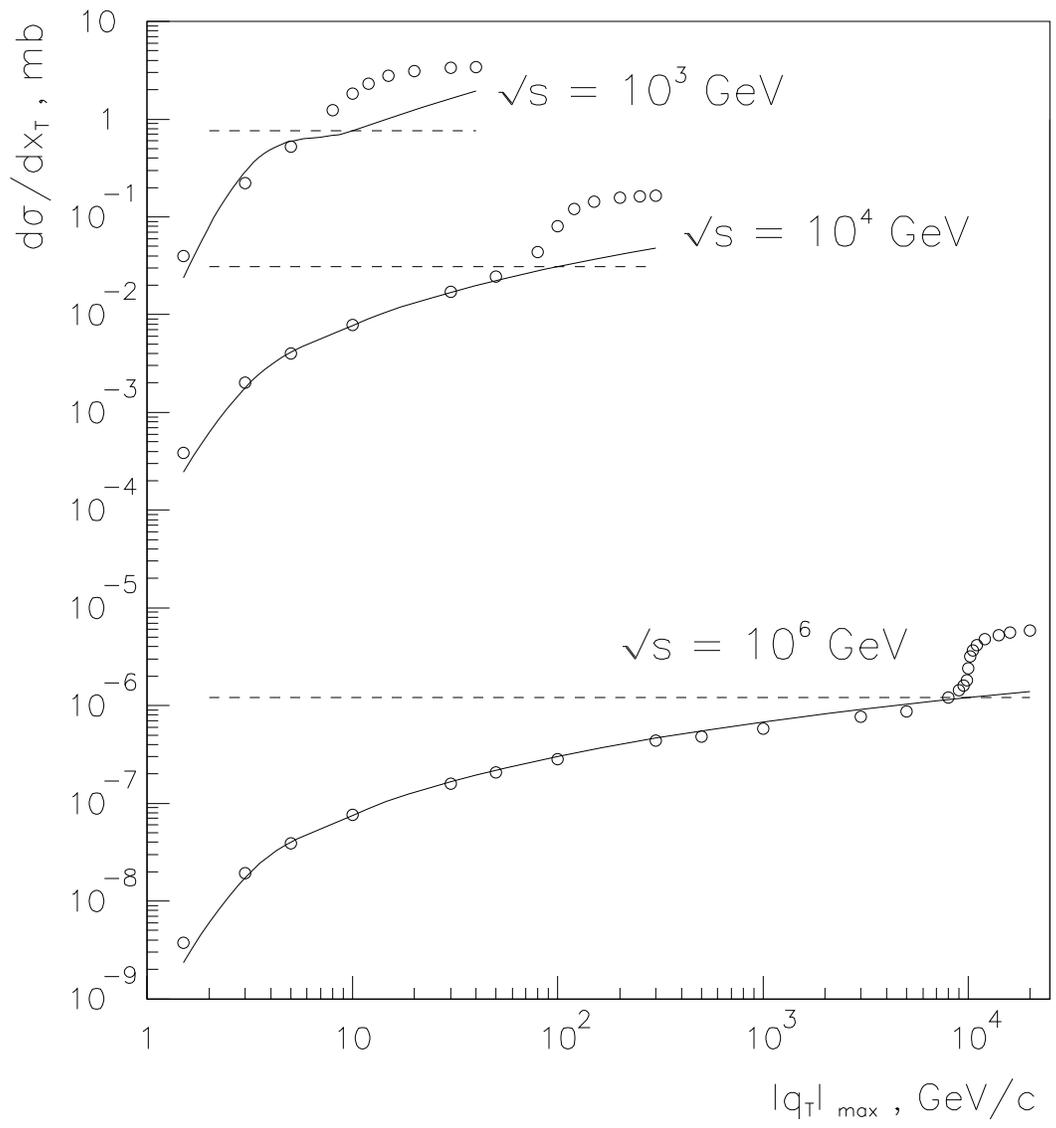
Fig. 3. The values of $d \sigma /d x_T$ for charm hadroproduction at
$x_T$ = 0.02 in $k_T$-factorization approach as a function of upper
limits of integrals in Eq.~(\ref{spp}) (points); the conventional parton
model values (dashed lines) and the values of right-hand side of
Eq.~(\ref{f}) (solid curves).

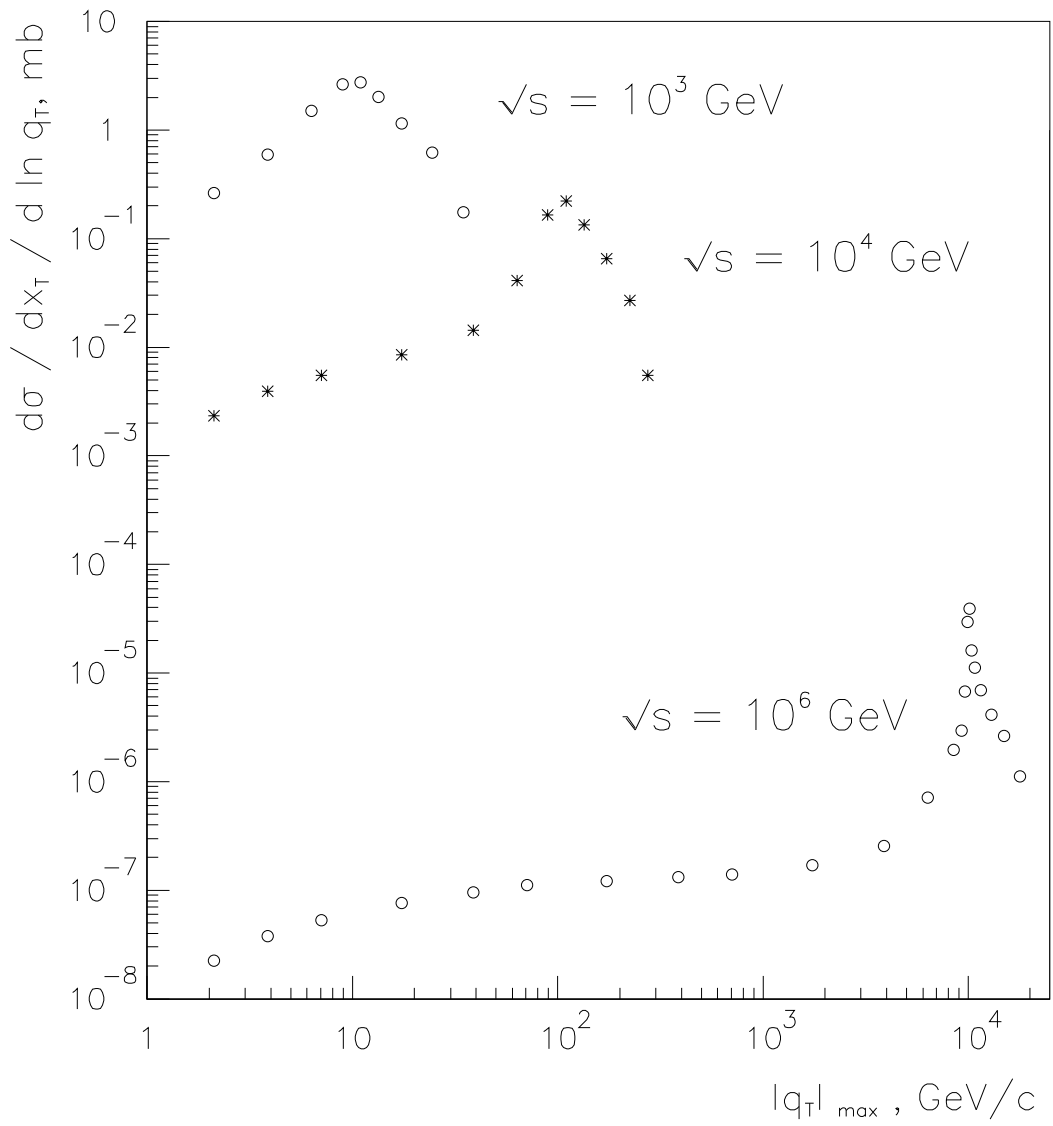
Fig. 4. The values of $d \sigma /d x_T/d \ln{q_{max}}$ for charm
hadroproduction at $x_T$ = 0.02 in $k_T$-factorization approach as a
function of upper limits of integrals in Eq.~(\ref{spp}) (points).

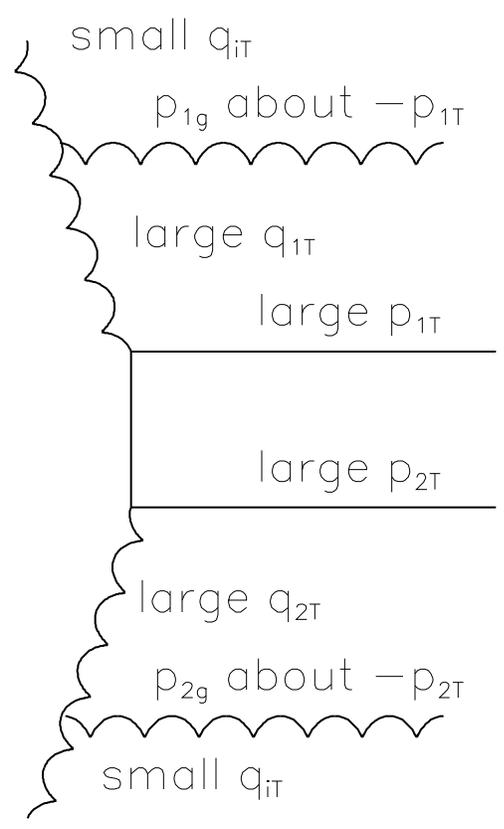
Fig. 5. The diagrams which are important in the case of one-particle
distributions of heavy quark with large $p_T$. The situation similar
to the LO parton model (a). The case, possible in NLO parton model,
large $p_T$ of the quark is compensated by hard gluon and the fermion
propagator is near to mass shell (b). The numerically small contribution, 
when large transverse momenta of heavy quarks are compensated by two hard 
gluons, whereas the fermion propagator is near to mass shell (c).

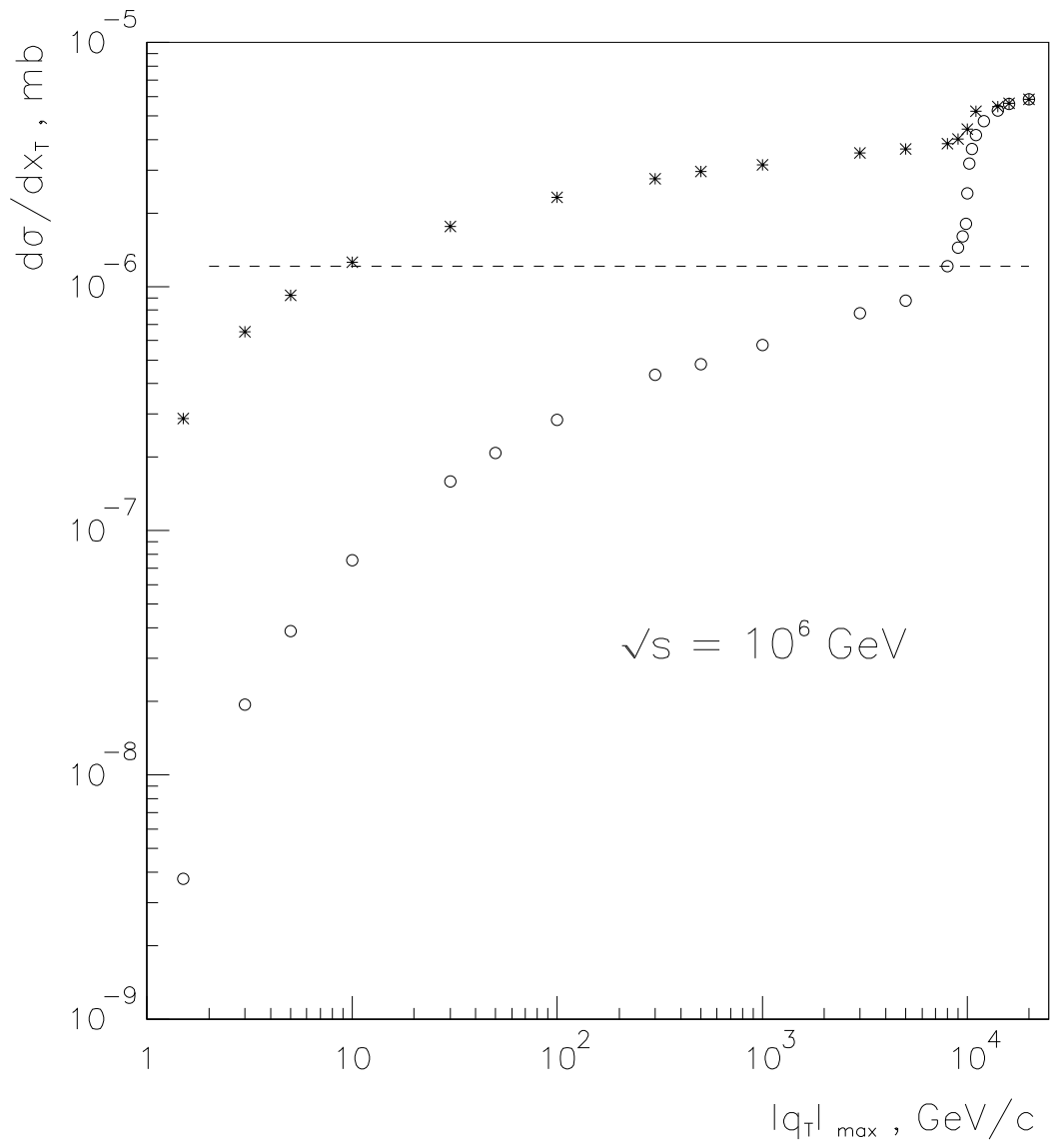
Fig. 6. The same as in Fig.~3 (points and dashed curves) together with
the values of $d \sigma /d x_T$ (stars) for the case when only $q_{2T}$
upper limit integration is bounded by $q_{max}$.

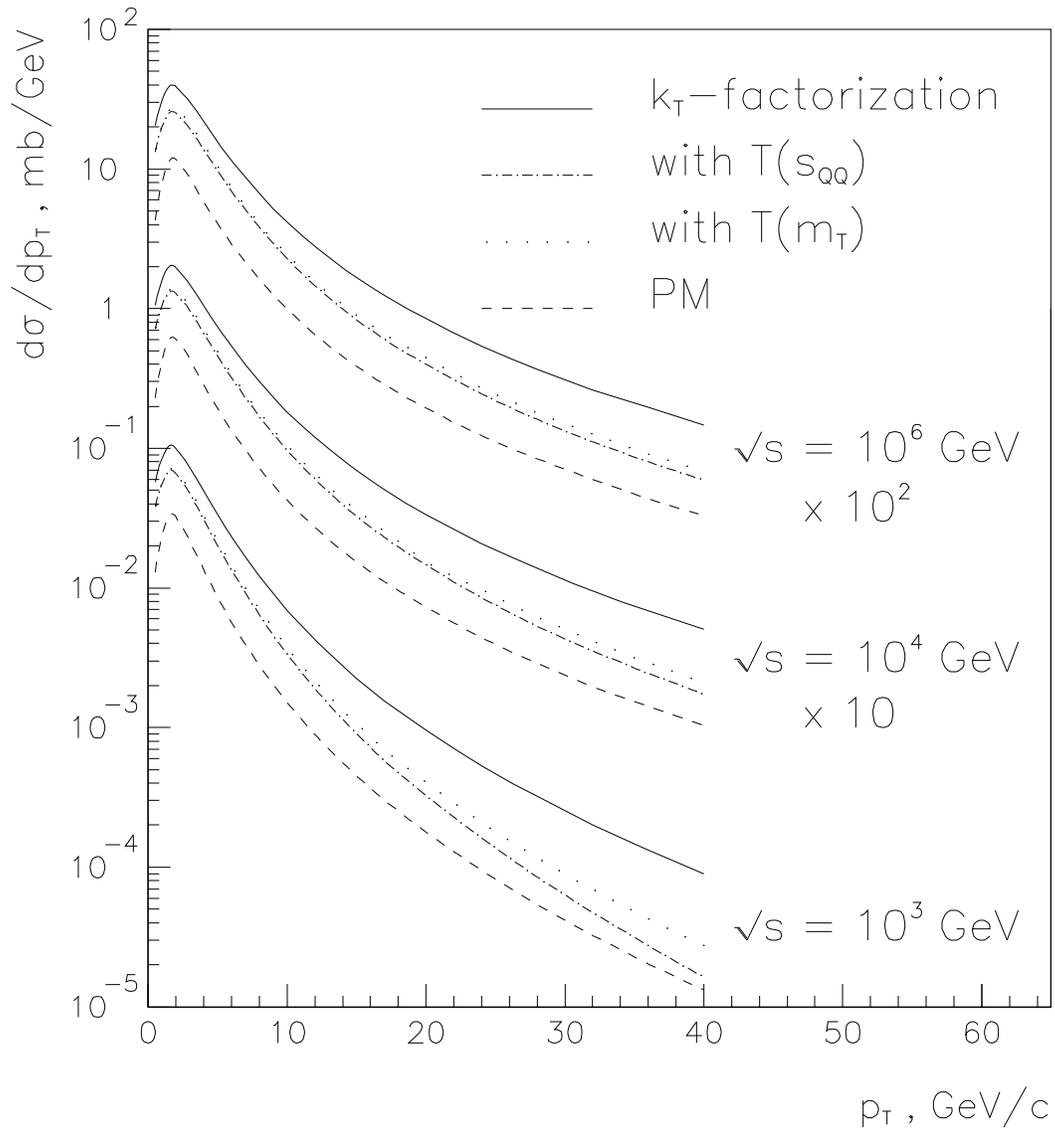
Fig. 7. $p_T$-distributions of $c$-quarks produced at different energies.
Dashed curves are the results of LO parton model. Solid curves are
calculated with unintegrated gluon distribution $\varphi(x,q^2)$ given by
Eq. (10), dash-dotted and dotted curves from Eqs. (11) and (12) for
$\mu^2$ values in Eq. (12) equal to $\hat{s}/4$ and $m_T^2$, respectively.

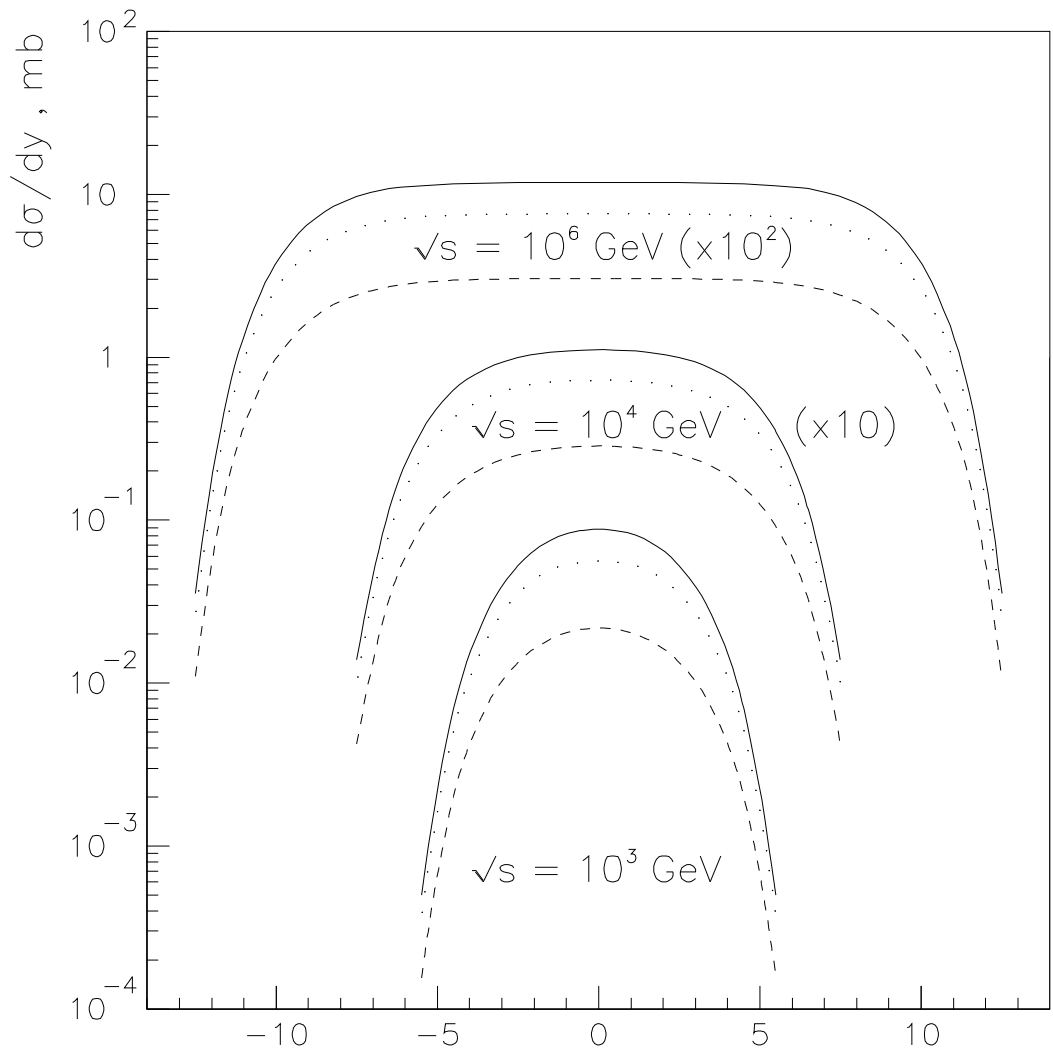
Fig. 8. Rapidity distributions of $c$-quarks produced at different energies.
Dashed curves are the results of LO parton model. Solid curves are
calculated with unintegrated gluon distribution $\varphi(x,q^2)$ given by
Eq. (10) and dotted curves from Eqs. (11) and (12) for
$\mu^2$ values in Eq. (12) equal to $m_T^2$.


\newpage

\begin{thebibliography}{99}

\bibitem{1} P.Nason, S.Dawson and R.K.Ellis. Nucl.Phys. B303 (1988) 607.
\bibitem{2} G.Altarelli et al. Nucl.Phys. B308 (1988) 724.
\bibitem{NDE} P.Nason, S.Dawson and R.K.Ellis. Nucl.Phys. B327 (1989)
49.
\bibitem{Beer} W.Beenakker, H.Kuijf, W.L.Van Neerven and J.Smith.
Phys.Rev. D40 (1989) 54.
\bibitem{Beer1} W.Beenakker, W.L.Van Neerven, R.Meng, G.A.Schuler
and J.Smith. Nucl.Phys. B351 (1991) 507.
\bibitem{BEAT} BEATRICE Coll. M.Adamovich et al. Phys.Lett. B348 (1995) 256.
\bibitem{BEAT1} BEATRICE Coll. Y.Alexandrov et al. Phys.Lett B433 (1998) 217.
\bibitem{MNR} M.N.Mangano, P.Nason and G.Ridolfi. Nucl. Phys. B373
(1992) 295.
\bibitem{FMNR} S.Frixione, M.N.Mangano, P.Nason and G.Ridolfi. Preprint
CERN-TH/97-16 (1997); hep-ph/9702287.
\bibitem{Shab} Yu.M.Shabelski. Talk, given at HERA Monte Carlo Workshop,
27-30 April 1998, DESY, Hamburg; hep-ph/9904492.
\bibitem{CCH} S.Catani, M.Ciafaloni and F.Hautmann. Phys.Lett. B242
(1990) 97; Nucl.Phys. B366 (1991) 135.
\bibitem{CE} J.C.Collins and R.K.Ellis. Nucl.Phys. B360 (1991) 3.
\bibitem{MW} G.Marchesini and B.R.Webber. Nucl.Phys. B386 (1992) 215.
\bibitem{CH} S.Catani and F.Hautmann. Phys.Lett. B315 (1993) 475;
Nucl.Phys. B427 (1994) 475.
\bibitem{CC} S.Camici and M.Ciafaloni. Nucl.Phys. B467 (1996) 25;
Phys.Lett. B396 (1997) 406.
\bibitem{GLR} L.V.Gribov, E.M.Levin and M.G.Ryskin. Phys.Rep. 100
(1983) 1.
\bibitem{LR} E.M.Levin and M.G.Ryskin. Phys.Rep. 189 (1990) 267.
\bibitem{8} E.M.Levin, M.G.Ryskin, Yu.M.Shabelski and A.G.Shuvaev.
Sov. J. Nucl. Phys. 53 (1991) 657.
\bibitem{lrs} E.M.Levin, M.G.Ryskin, Yu.M.Shabelski and A.G.Shuvaev.
Sov.J.Nucl.Phys. 54 (1991) 867.
\bibitem{DDT} Yu.L.Dokshitzer, D.I.Dyakonov and S.I.Troyan.
Phys.Rep. 58 (1980) 270.
\bibitem{ASS} Ya.I.Azimov, Yu.M.Shabelski and O.P.Strogova. Phys.Atom.
Nucl. 57 (1994) 674.
\bibitem{CSS} J.C.Collins, D.E.Soper and G.Sterman. Nucl.Phys. B308
(1988) 833.
\bibitem{MNR1} M.N.Mangano, P.Nason and G.Ridolfi. Nucl. Phys. B405
(1993) 507.
\bibitem{Blu} J.Bl\"{u}mlein. Preprint DESY 95-121 (1995).
\bibitem{BFKL} E.A.Kuraev, L.N.Lipatov and V.S.Fadin. Sov. Phys. JETP 45
(1977) 199.
\bibitem{KS} G.P.Korchemsky and G.Sterman. Nucl.Phys. B437 (1995) 415.
\bibitem{blm} S.J.Brodsky, G.P.Lepage and P.B.Mackenzie. Phys.Rev. D28
(1983) 228.
\bibitem{3} M.G.Ryskin, Yu.M.Shabelski and A.G.Shuvaev. Z.Phys. C69
(1996) 269.
\bibitem{Nar} S.Narison. Phys.Lett. B341 (1994) 73.
\bibitem{BBB} P.Ball, M.Beneke and V.M.Braun. Phys.Rev. D52 (1995) 3929.
\bibitem{Liu} Liu Wenjie, O.P.Strogova, L.Cifarelli and
Yu.M.Shabelski. Phys.Atom.Nucl. 57 (1994) 844.
\bibitem{Kwi} J.Kwiechinski. Z.Phys. C29 (1985) 561.

\end{thebibliography}
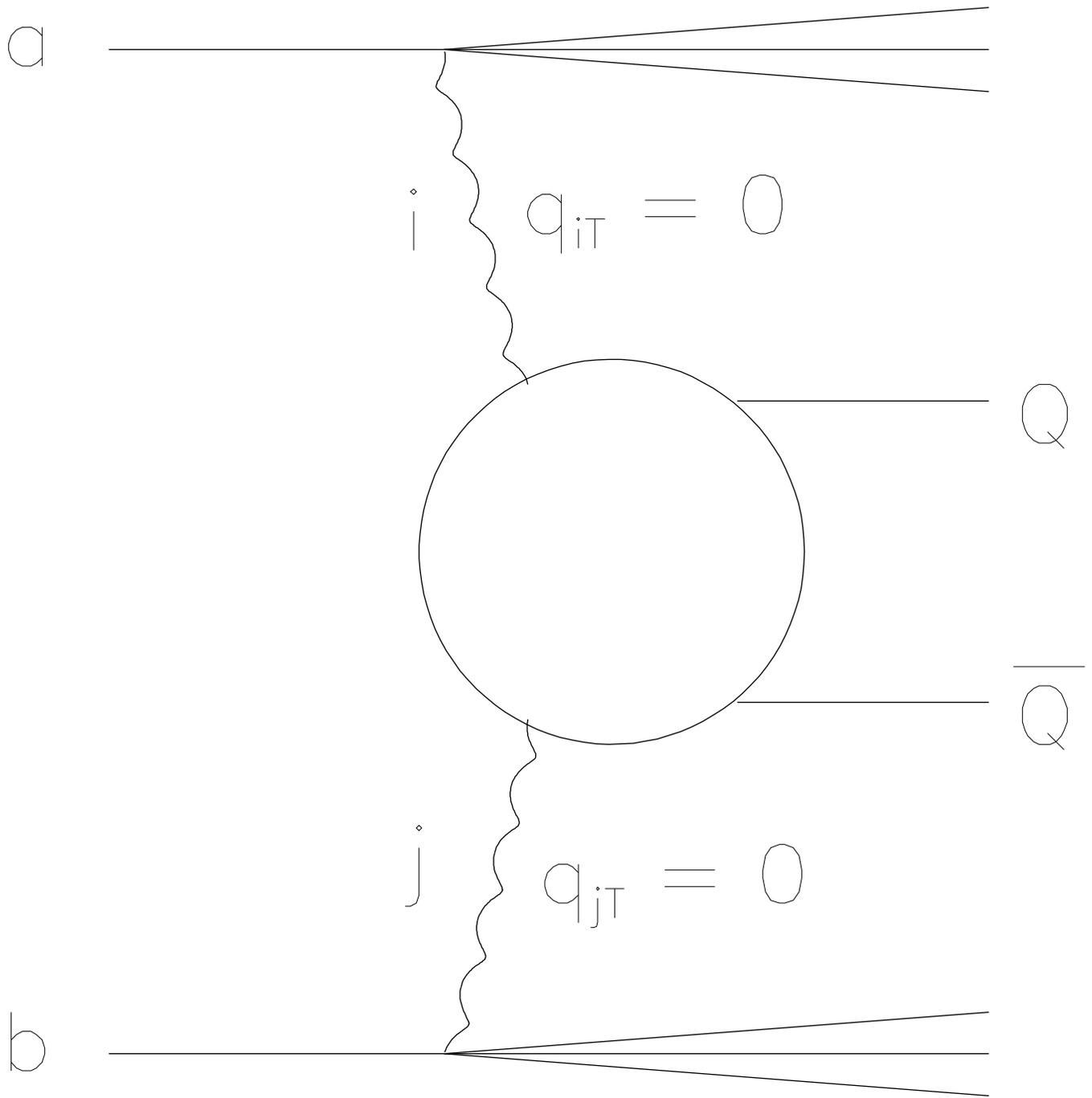
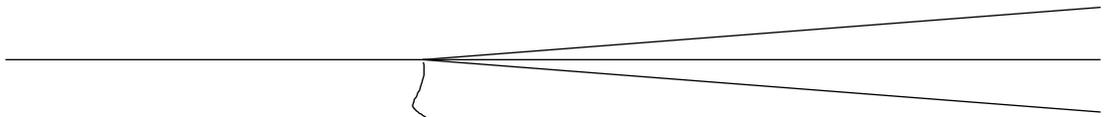
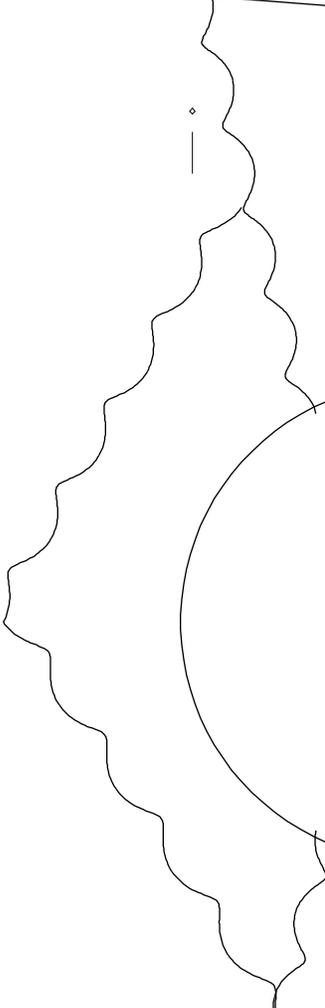
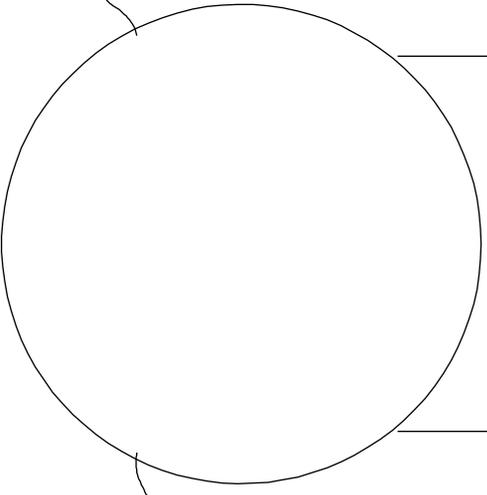
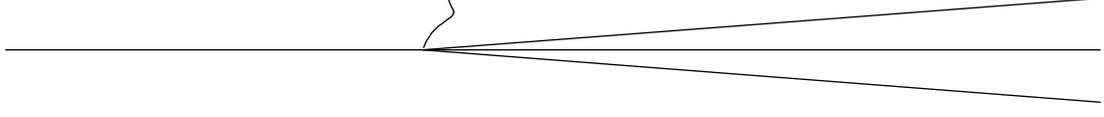
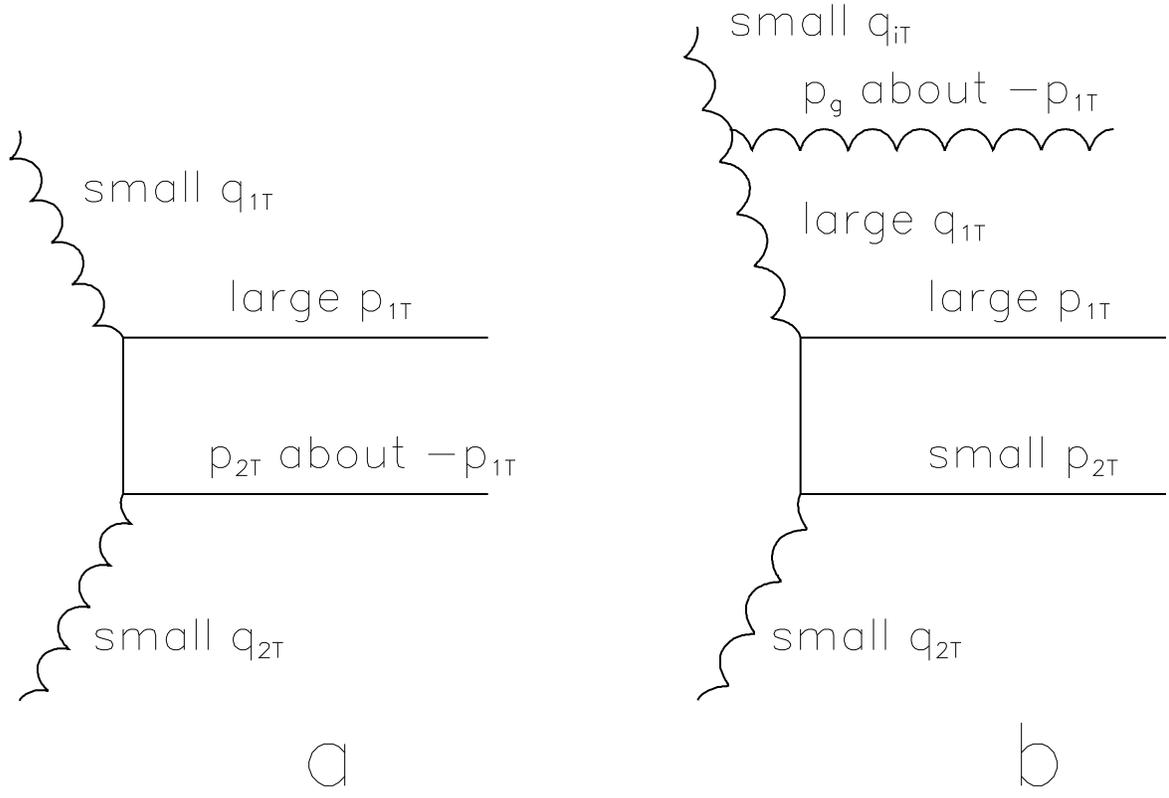
\end{document}